\documentclass[a4paper,onecolumn,11pt,accepted=2020-09-19]{quantumarticle}
\pdfoutput=1
\usepackage[utf8]{inputenc}
\usepackage[english]{babel}
\usepackage[T1]{fontenc}
\usepackage{amsmath,amssymb}
\usepackage{hyperref}
\usepackage[sort&compress,numbers]{natbib}
\usepackage{tikz}
\usepackage{lipsum}

\newcommand{\bra}[1]{\langle {#1} \vert}
\newcommand{\ket}[1]{\vert {#1} \rangle}

\newcommand{\del}[0]{\partial}

\newtheorem{proposition}{Proposition}
\begin{document}

\title{Distillation of maximally correlated bosonic matter from many-body quantum coherence}

\author{T.\,J.\,Volkoff}
\affiliation{Theoretical Division, Los Alamos National Laboratory, Los Alamos, NM 87545, USA}
\orcid{0000-0002-5511-3913}

\maketitle

\begin{abstract}
 We construct quantum coherence resource theories in symmetrized Fock space (QCRTF), thereby providing an information-theoretic framework that connects analyses of quantum coherence in discrete-variable (DV) and continuous variable (CV) bosonic systems. Unlike traditional quantum coherence resource theories, QCRTF can be made independent of the single-particle basis and allow to quantify coherence within and between particle number sectors. For example, QCRTF can be formulated in such a way that neither Bose-Einstein condensates nor Heisenberg-Weyl coherent states are considered as quantum many-body coherence resources, whereas spin-squeezed and quadrature squeezed states are.  The QCRTF framework is utilized to calculate the optimal asymptotic distillation rate of maximally correlated bosonic states both for particle number conserving resource states and resource states of indefinite particle number. In particular, we  show how to generate a uniform superposition of maximally correlated bosonic states from a state of maximal bosonic coherence with asymptotically unit efficiency using only free operations in the QCRTF.
\end{abstract}

\section{Introduction}

Protocols for transferring quantum coherence from discrete variable (DV) quantum systems to continuous variable (CV) quantum systems, and \textit{vice versa}, provide vital links between quantum information processing platforms \cite{furusawarev}. Although most proposed protocols involve explicit models for light-matter dynamics \cite{PhysRevB.94.205118,zeng}, for analyses of fundamental limits on processing quantum coherence it is useful to restrict the availability of quantum states and channels according to a resource-theoretic framework \cite{PhysRevLett.116.080402,PhysRevA.94.052324,pleniocoherence}. For example, utilizing local Gaussian operations (i.e., quantum channels that locally map Gaussian states to Gaussian states), it is possible to produce Gaussian entanglement from non-Gaussian CV quantum states (which may be obtained from a hybrid DV-CV state preparation protocol) \cite{pleniobrowne}. The recent development of quantum coherence resource theories (QCRT) for DV and CV quantum systems allow to place such protocols within the more general context of coherence manipulations \cite{pleniocoherence,tanvolk,PhysRevA.93.012334,PhysRevA.93.032111}.

In previously studied QCRTs, the free states of composite quantum systems are constructed from the tensor product of free states of a single subsystem \cite{adessoconvert,winteryang,PhysRevA.92.022112,PhysRevLett.116.150504}. When the free states of a single subsystem are separable, this necessarily implies that free states of the composite system are separable. However, in the case of bosonic matter, all states that are not statistical mixtures of Bose-Einstein condensates (BEC) exhibit entanglement between particles. This fact motivates the analysis of QCRTs for identical particles \cite{PhysRevA.96.032334,PhysRevLett.112.150501,PhysRevB.98.014519} and suggests the possibility of considering statistical mixtures of $SU(M)$ coherent states as free states in a QCRT.

In the present work, we take this idea further to define three interrelated \textit{quantum coherence resource theories on bosonic Fock space} (QCRTF) that provide a framework to analyze quantum coherence in DV and CV bosonic quantum systems. These QCRTF exhibit fundamental differences from traditional QCRTs. For example, the free states in the QCRTFs defined in this work do not necessarily have fixed particle number, and two of the QCRTFs are independent of the single particle basis. Section \ref{sec:summ} gives an overview of the physical intuition behind the sets of free states and free operations in the three versions of QCRTF that are analyzed in detail in Sections \ref{sec:aa}, \ref{sec:bb}, \ref{sec:cc}, respectively. In Section \ref{sec:physconsid} we discuss relevant coherence measures and some features of QCRTF that are relevant to practical implementations, including the possibility of constraining the set of free quantum channels to preserve the energy density according to a partition of the single-particle modes. The main results in Section \ref{sec:ex} consist of calculations of the rate of distillation of maximal correlations between pairs of modes (i.e., determination of the largest $R>0$ such that, as $n\rightarrow \infty$, $nR$ copies of a two-mode maximally correlated (MC) state with energy $E$ are produced from $n$ copies of a given two-mode quantum state with energy $E$ by using only free quantum channels in the QCRTF). This approach yields the asymptotic distillation rate of maximally correlated states from Bose-Einstein condensates, pair correlated states \cite{PhysRevA.96.023621}, and states of indefinite particle number possessing maximal coherence. In the latter case, we show that preparation of maximal coherence with respect to QCRTF allows extraction of intermode correlations with asymptotically unit efficiency.

The setting of our analysis is the symmetrized Fock space $\mathcal{F}_{M}=\bigoplus_{N=0}^{\infty}\mathcal{F}^{(N)}_{M}$ associated to a $M$-dimensional single-particle Hilbert space $\mathcal{H}$, where $\mathcal{F}^{(N)}_{M}:= S_{N}(\mathcal{H})$ is the symmetric subspace of the tensor product $\mathcal{H}^{\otimes N}$ \cite{symmchurch}, and $S_{0}(\mathcal{H})$ is the trivially symmetric Hilbert space $\mathbb{C}$ (which can be considered the vacuum sector $\mathbb{C}\ket{\text{VAC}}$). The symbol $P_{B}$ denotes the direct sum (over $N$) of projections from $\mathcal{H}^{\otimes N}$ to $S_{N}(\mathcal{H})$ ($P_{B}$ is also called the Bose projector). The $N$-particle sector $\mathcal{F}^{(N)}_{M}$ can be related to the symmetrized Fock space $\mathcal{F}_{M}$ by writing $\mathcal{F}^{(N)}_{M}:= Q_{N}\mathcal{F}_{M}$ where $Q_{N}$ is the $N$-particle sector projection. For a physical picture of the Hilbert space $\mathcal{F}_{M}$, one can consider, e.g., optically trapped ultracold bosons in the grand canonical ensemble. We use the generic notation $D(\mathcal{K})$ for quantum states on a given Hilbert space $\mathcal{K}$. An asterisk denotes the adjoint of a linear operator, $N_{M}:=\sum_{j=1}^{M}a_{j}^{*}a_{j}$ denotes the particle number operator, and $\mathcal{N}_{M}(\rho):=\text{tr}N_{M}\rho$ denotes the particle number functional.

\section{\label{sec:summ}Summary and organization}
Ananlogously to the construction of QCRT for DV systems, a QCRTF is defined by the following data: 1. a closed, convex set of \textit{free states} contained in $D(\mathcal{F}_{M})$, 2. a set of \textit{free operations}, which are quantum channels defined relative to the set of free states, and 3. a faithful coherence measure satisfying a monotonicity property under free operations \cite{pleniocoherence}.
The present work introduces three versions of QCRTF, called QCRTF$^{(A,B,C)}$, that are respectively associated with sets of free states $\Delta^{(A,B,C)}_{M}$ on $M$ single-particle modes that satisfy $\Delta^{A}_{M}\subset \Delta^{B}_{M} \subset \Delta^{C}_{M}$. Further, states in $\Delta^{B}_{M}$ can exhibit quantum coherence between pure states of $\Delta^{A}_{M}$, and states in $\Delta^{C}_{M}$ can exhibit quantum coherence between pure states of $\Delta^{B}_{M}$.  The set $\Delta^{A}_{M}$ can be pictured as statistical mixtures of pure states having fixed local particle number with respect to a set of $M$ single particle modes. For example, a statistical mixture of insulator states (i.e., Fock states) or a pure Bose-Einstein condensate in one of the $M$ single particle modes are in $\Delta^{A}_{M}$. 

The set $\Delta^{B}_{M}$ can be pictured as a mixture of number-conserving linear optical quantum computations acting on $\Delta^{A}_{M}$. Although such states are, in fact, a strict subset of $\Delta^{B}_{M}$, this picture allows one to conclude that statistical mixtures of Bose-Einstein condensates in different single-particle modes, and the images of networks of beamsplitters on Fock states \cite{aaronson} are in $\Delta^{B}_{M}$. Finally, the set $\Delta^{C}_{M}$ includes states of indefinite particle number by allowing mixtures of quantum optical displacement operators to be applied to $\Delta^{B}_{M}$. Therefore, $\Delta^{C}_{M}$ include Heisenberg-Weyl coherent states (in fact, all classical states in quantum optics), and displaced Fock states. Section \ref{sec:aa}, \ref{sec:bb}, \ref{sec:cc} focus respectively on QCRTF$^{(A,B,C)}$.

The sets of free operations $\mathcal{E}^{(A,B,C)}_{M}$ are CV analogues of a set of free operations called IO in the DV quantum coherence resource theory \cite{winteryang}. The motivation for IO (and, therefore, for $\mathcal{E}^{(A,B,C)}_{M}$) is that a free operation with Kraus operators $\lbrace K_{\beta} \rbrace_{\beta}$ should not generate coherence from a free state $\rho$ even if the state proportional to $K_{\beta}\rho K_{\beta}^{*}$ can be selected via, e.g., measurement of an environment. However, these channels are not \textit{physically free} in the sense that they can not be defined by appending of a free environment state, followed application of a free unitary, followed by discarding quantum information. In the respective subsections, we discuss how subsets of physically free quantum channels can be included in $\mathcal{E}^{(A,B,C)}_{M}$. Finally, in Section \ref{sec:ex}, we consider the operational import of QCRTF$^{(A,B,C)}$ in terms of the task of distilling maximally correlated (i.e., maximally mode entangled \cite{PhysRevLett.91.097902}) bosonic states using only operations from $\mathcal{E}^{A}_{M}$. In such a task, coherence with respect to QCRTF$^{(A)}$ is valuable, so one can consider input states consisting of bosonic copies of states from $\Delta_{2}^{(B,C)}$ (Section \ref{sec:nc}) or, e.g., maximally coherent states with respect to QCRTF$^{(A)}$ (Section \ref{sec:nonnumcons}). In the latter scenario, we find that maximally correlated states can be distilled with asymptotically unit efficiency.

\section{\label{sec:fockcoh}Quantum coherence in bosonic Fock space}

We motivate and define three versions of QCRTF which have free state sets that are related by inclusion. However, the corresponding sets of free operations are not so simply related, and thus allow to analyze a wide variety of information processing protocols.

\subsection{\label{sec:aa}QCRTF$^{(A)}$}

As a first step, consider the restricted quantum coherence resource theory QCRT$^{(N)}_{M}$, which is associated to a particle number sector $\mathcal{F}^{(N)}_{M}$, $N\in \mathbb{Z}_{\ge 0}$.  Given a set of single particle modes $\lbrace \ket{1},\ldots,\ket{M}\rbrace$, the free states in QCRT$^{(N)}_{M}$ are defined as the discrete probability measures on the set of Fock states $\ket{\vec{m}}\bra{\vec{m}}$ (\textit{i.e.}, fixed particle number states \footnote{A Fock state $\ket{\vec{m}}$ on orthonormal modes $\ket{\psi_{1}}, \ldots,\ket{\psi_{M}}$ is defined by $\ket{\vec{m}}\propto \prod_{j=1}^{M}a_{\psi_{j}}^{*m_{j}}\ket{\text{VAC}}$. See Ref.\cite{robinsonvol2} for the algebra of the canonical commutation relations and its action on the symmetrized Fock space.}) where $\vec{m}=(m_{1},\ldots , m_{M}) \in (\mathbb{Z}_{\ge 0})^{\times M}$ and $\vert \vec{m} \vert:= \sum_{j=1}^{M}m_{j}=N$. We refer to this set of free states as $\Delta^{(N)}_{M}$, its extreme points (pure free states) as $\mathfrak{P}^{(N)}_{M}$ and take $\Delta^{(0)}_{M}:=\mathfrak{P}^{(0)}_{M}= \ket{\text{VAC}}\bra{\text{VAC}}$, the empty vacuum, for all $M$.  The set of free operations $\mathcal{E}^{(N)}_{M}$ is taken to be those completely positive, trace perserving (CPTP) quantum channels having Kraus form $\Phi(\rho)=\sum_{\beta}K_{\beta}\rho K_{\beta}^{*}$ where the $K_{\beta}$ are bounded operators on $\mathcal{F}^{(N)}_{M}$ and further satisfy \begin{eqnarray}&{}&1.\;\sum_{\beta}K_{\beta}^{*}K_{\beta}=\mathbb{I}^{(N)} \nonumber \\ &{}&2.\; K_{\beta}\rho K_{\beta}^{*}/\text{tr}K_{\beta}\rho K_{\beta}^{*}\in\Delta_{M}^{(N)}\; \forall \; \rho \in \Delta^{(N)}_{M}\label{eqn:freeopA}\end{eqnarray} for all indices $\beta$. $\mathcal{E}^{(N)}_{M}$ is a proper subset of the set of CPTP $\Phi$ such that $\Phi(\Delta^{(N)}_{M})\subset \Delta^{(N)}_{M}$, except in the cases of QCRT$^{(0)}_{M}$ and QCRT$^{(N)}_{1}$, in which the respective identity channels are the only free channels. Note that due to the bosonic symmetry, QCRT$^{(N)}_{M}$ is not equivalent to a tensor product of quantum coherence resource theories of a finite dimensional Hilbert space with orthonormal basis $\lbrace \ket{1},\ldots , \ket{M}\rbrace$ defining the free states \cite{PhysRevLett.116.150504}. For example, in such a tensor product, the two-particle, non-bosonic state $\ket{1}_{A}\otimes \ket{2}_{B}$ is free, where $A$ and $B$ label the two copies of the Hilbert space.  

With the definition of the QCRT$_{M}^{(N)}$ in hand, we can now define a quantum coherence resource theory QCRTF$^{(A)}$ that takes into account the different particle number sectors. QCRTF$^{(A)}$ is defined by an orthonormal basis $\lbrace \ket{1},\ldots ,\ket{M}\rbrace$ and a set of free states $\Delta_{M}^{A}$, defined as follows: $\rho \in \Delta^{A}_{M}$ if and only if for any $0<\epsilon < 2$, there is an $N\in\mathbb{Z}_{\ge 0}$, a probability density $p:  \bigoplus_{j=0}^{N} \mathfrak{P}_{M}^{(j)} \rightarrow \mathbb{R}$, and $\rho'$ having the form
\begin{eqnarray}
\rho ' &=&  \bigoplus_{j=0}^{N}\sum_{\sigma^{(j)}} p(\sigma^{(j)})\sigma^{(j)}
\label{eqn:freeA}
\end{eqnarray}
such that $\Vert \rho - \rho ' \Vert_{1} < \epsilon$. In (\ref{eqn:freeA}), the sum over $\sigma^{(j)}$ runs over a countable set of states in $\mathfrak{P}^{(j)}_{M}$ and the probability density is normalized as $\sum_{j=0}^{\infty}\sum_{\sigma^{(j)}}p(\sigma^{(j)})=1$. The set of free operations is defined as $\mathcal{E}_{M}^{A}:= \oplus_{N\ge 0}\mathcal{E}^{(N)}_{M}$. As a concrete example of states in $\Delta_{M}^{A}$, we note that for $L\in \mathbb{N}$, the set $\Delta_{M=L^{d}}^{A}$ contains Gibbs states of massive, non-interacting bosons on the lattice $\lbrace 1,\ldots ,L \rbrace^{\times d}$ at temperature $T\ge 0$ and chemical potential $\mu \in \mathbb{R}$. Free quantum channels from $\Delta^{A}_{M}$ to $\Delta^{A}_{M'}$ for $M'\neq M$ are constructed via $\Phi_{2}\circ \mathcal{Q} \circ \Phi_{1}$ where $\Phi_{1(2)} \in \mathcal{E}_{M(M')}^{A}$, and $\mathcal{Q}$ is the partial trace over modes (if $M'<M$) or bosonic mode appending channel (if $M'>M$).  Bosonic modes are appended by preparation of a free state followed by Bose symmetrization. Specifically, if $\lbrace \ket{M+1},\ldots ,\ket{M'}\rbrace$  is orthonormal to $\lbrace \ket{1},\ldots ,\ket{M}\rbrace$ and $\ket{\vec{n}}$ is a Fock state on modes $\lbrace \ket{M+1},\ldots ,\ket{M'}\rbrace$ with $\vert \vec{n}\vert=N'$ (i.e., it is an element of $\mathfrak{P}^{(N')}_{M'-M}$ with respect to those modes), then $\ket{\vec{n}}$ defines the mode appending channel acting as $\ket{\vec{m}} \rightarrow \ket{\vec{m},\vec{n}}\in \mathfrak{P}_{M'}^{(\vert \vec{m}\vert+N')}$. This channel maps $\Delta^{A}_{M}$ to $\oplus_{N\ge N'}\Delta^{(N)}_{M'} \subset \Delta^{A}_{M'}$.

It is important to note that not every element of the set of free operations in QCRTF$^{(A)}$ given in (\ref{eqn:freeopA}) can be written as a concatenation of the following three quantum channels: appending a free environment state (as described above), evolving by a free system-plus-environment interaction, then carrying out a free projective measurement channel on the environment degrees of freedom. The concatenation of these channels comprises the set of \textit{physically free operations} (sometimes called PIO) \cite{chitambarpio}. Because $\Delta_{M}^{(N)}$ has a natural action by the permutation group on $M$ letters, the subset of $\mathcal{E}_{M}^{A}$ corresponding to physically free operations can be constructed following Proposition 1 of Ref. \cite{chitambarpio}. In Section \ref{sec:bb}, we will construct a resource theory QCRTF$^{(B)}$ associated with a set of free operations that excludes a large class of physically motivated channels, so a more careful consideration of physically free operations is required.

\subsection{\label{sec:bb}QCRTF$^{(B)}$}
In the examples that we consider in Section \ref{sec:ex}, it will be useful to define a more general QCRTF having free states consisting of statistical mixtures of states of different particle number as in QCRTF$^{(A)}$, but is independent of the single-particle basis. In order to do this, we review the set of \textit{linear optical unitary channels} which is labeled $\mathcal{U}_{M}$. Given $M$ orthonormal modes $\ket{1},\ldots, \ket{M}$ spanning Hilbert space $\mathcal{H}$, one first constructs the vector of canonical operators $R=(\hat{q}_{1},\hat{p}_{1},\ldots ,\hat{q}_{M},\hat{p}_{M})$, where $\hat{q}_{j}:= 2^{-1/2}(a_{j}+a_{j}^{*})$, $\hat{p}_{j}=2^{-1/2}(-ia_{j}+ia_{j}^{*})$. The set $\mathcal{U}_{M}$ is defined as those unitary channels $\rho \mapsto U\rho U^{*}$, where the unitary $U$ acts on the vector of canonical operators  via $U^{*}RU=RT + b$ for an orthogonal matrix $T\in SO(2M,\mathbb{R})\cap Sp(2M,\mathbb{R})\cong U(M)$ and $b\in \mathbb{R}^{2M}$ \cite{serafini}.  It is also useful to define the subset $\tilde{\mathcal{U}}_{M} \subset \mathcal{U}_{M}$ of linear optical unitaries that commute with the number operator $N_{M}:=\sum_{j=1}^{M}a_{j}^{*}a_{j}$ (and, therefore, the unitaries corresponding to elements of $\tilde{\mathcal{U}}_{M}$ have $b=0$ in their linear action on the canonical variables). We will always consider the action of $\mathcal{U}_{M}$ or $\tilde{\mathcal{U}}_{M}$ in symmetrized Fock space.  
A phase space displacement operator is an element  $\mathcal{U}_{M}$ with $T$ given by the $2M\times 2M$ identity matrix and $b\neq 0$. We write a displacement operator corresponding to $\alpha \in \mathbb{C}$ as $D(\alpha)=e^{\alpha a_{\psi}^{*}-\overline{\alpha}a_{\psi}}$ for $\ket{\psi}\in\mathcal{H}$ so that a Heisenberg-Weyl coherent state $e^{\alpha a_{\psi}^{*}-\overline{\alpha}a_{\psi}}\ket{\text{VAC}}$ is an element of the symmetric Fock space with projection on the $N$ particle sector given by $e^{-\vert \alpha \vert^{2}/2}{(\alpha a_{\psi}^{*})^{N}\over N!}\ket{\text{VAC}}$.

The quantum coherence resource theory QCRTF$^{(B)}$ is defined by first considering the bosonic Fock space $\mathcal{F}_{M}$ with respect to $\mathcal{H}$ and the set of number-conserving linear optical unitary channels $\tilde{\mathcal{U}}_{M}$. 
For QCRTF$^{(B)}$, the set of free states $\Delta^{B}_{M}$ is defined as follows: $\rho \in \Delta^{B}_{M}$ if and only if for any $0<\epsilon < 2$, there is an $N\in\mathbb{Z}_{\ge 0}$, a joint probability density $p:  \bigoplus_{j=0}^{N}\left( \mathfrak{P}_{M}^{(j)} \times  U(M) \right) \rightarrow \mathbb{R}$, and $\rho'$ having the form
\begin{eqnarray}
\rho ' &=&  \bigoplus_{j=0}^{N}\sum_{\sigma^{(j)}} \int \mu(dg_{j})\,p(g_{j},\sigma^{(j)})U_{g_{j}}\sigma^{(j)} U_{g_{j}}^{*}
\label{eqn:freeB}
\end{eqnarray}
such that $\Vert \rho - \rho ' \Vert_{1} < \epsilon$. In (\ref{eqn:freeB}), the sum over $\sigma^{(j)}$ runs over a countable set of states in $\mathfrak{P}^{(j)}_{M}$, the probability density is normalized as $\sum_{j=0}^{\infty}\sum_{\sigma^{(j)}}\int \mu(dg_{j})p(g_{j},\sigma^{(j)})=1$, $\mu(dg_{j})$ is the normalized Haar measure on $U(M)$ for $j>1$ and a point measure at the identity matrix for $j=0$ \cite{parthasarathy,Braun_2006}. For a physical interpretation of (\ref{eqn:freeB}), it is useful to consider the subset of $\Delta^{B}_{M}$ that is obtained by applying statistical mixtures of linear optical quantum computations to $\Delta^{A}_{M}$. This subset corresponds to the subset of free states defined by probability densities $p(g_{j},\sigma_{j})$ that are factorizable in $g_{j}$ and $\sigma_{j}$ arguments. Applying the quantum channel $\rho \mapsto \lim_{N\rightarrow \infty}\sum_{\sigma \in \mathfrak{P}^{(0)}_{M} \oplus \cdots \oplus \mathfrak{P}^{(N)}_{M}} \sigma \rho \sigma $ to $\Delta^{B}_{M}$ gives $\Delta_{M}^{A}$. 

The free set $\Delta^{B}_{M}$ has striking features which makes it appropriate for the consideration of many-body quantum coherence, as opposed to single-particle quantum coherence. For example, unlike QCRT$^{(1)}_{M}$, all single-particle quantum states (i.e., elements of $\mathcal{H}$) are contained in $\Delta^{B}_{M}$. By the same reasoning, statistical mixtures of Bose-Einstein condensates in different single-particle modes are contained in $\Delta^{B}_{M}$. Certain $N$-particle pure states of $\Delta_{M}^{B}$, namely those corresponding to a probability density $p(g_{j},\sigma^{(j)})$ which is proportional to $\delta_{g_{j},g}$ for all $j$, can be expressed as the unique ground state of a twisted Bose-Hubbard interaction of the form
\begin{equation}H_{g}:= \sum_{i=1}^{M}U_{g}(a_{i}^{*}a_{i}-m_{i})^{2}U_{g}^{*}\label{eqn:twisted}\end{equation} 
where $\sum_{i=1}^{M}m_{i}=N$. The interaction (\ref{eqn:twisted}) can be used to describe locally varying potentials such as those that arise in the physics of ultracold atoms in random optical traps \cite{FALLANI2008119}. 

We now define the set of free operations for QCRTF$^{(B)}$. This set, labeled $\mathcal{E}_{M}^{B}$, is taken to consist of quantum channels $\Phi$ having Kraus decomposition is given by $\Phi(\rho)=\sum_{\beta}K_{\beta}\rho  K_{\beta}^{*}$, where $K_{\beta}:\mathcal{F}_{M}\rightarrow \mathcal{F}_{M}$ are bounded and further satisfy
\begin{eqnarray}
{}&{}&\text{1. }\sum_{\beta}K_{\beta}^{*}K_{\beta} = \mathbb{I}_{\mathcal{F}_{M}}\nonumber \\
{}&{}&\text{2. }K_{\beta}\rho  K_{\beta}^{*} /\text{tr}K_{\beta}\rho  K_{\beta}^{*}\in \Delta_{M}^{B} \;\; \forall \; \rho \in \Delta_{M}^{B}
\label{eqn:embset}
\end{eqnarray} for all $\beta$.  Despite the proper subset relation of free states given by $\Delta_{M}^{A} \subset \Delta_{M}^{B}$, the set $\mathcal{E}_{M}^{A}$ is not contained in $\mathcal{E}_{M}^{B}$. For example, the projection-valued measurement channel
\begin{equation}
\rho \mapsto (\mathbb{I}_{\mathcal{F}_{2}} - P) \rho (\mathbb{I}_{\mathcal{F}_{2}} - P) + P\rho P  \label{eqn:measchan}\end{equation} where $P:= \ket{0,N}\bra{0,N} + \ket{N,0}\bra{N,0}$, is an element of $\mathcal{E}_{2}^{A}$ but not an element of $\mathcal{E}_{2}^{B}$ since, e.g., it contains a Kraus operator that maps the Bose-Einstein condensate $\propto (a_{1}^{*}+a_{2}^{*})^{N}\ket{\text{VAC}} \in \Delta_{2}^{B}$ to the GHZ state (NOON state) $(a_{1}^{*N}+a_{2}^{*N})\ket{\text{VAC}}$, which is not in  $\Delta_{2}^{B}$ if $N>2$. 
Finally, in the same way as we defined free mode appending quantum channels that map $\Delta^{A}_{M}$ to $\Delta^{A}_{M'}$, we can define mode appending quantum channels of the form $\Phi_{2}\circ \mathcal{Q} \circ \Phi_{1}$, where $\Phi_{1(2)} \in \mathcal{E}_{M(M')}^{B}$ with $M'> M$, that map $\Delta^{B}_{M}$ to $\Delta^{B}_{M'}$. However, discarding quantum information is not always free with respect to QCRTF$^{(B)}$, as we discuss presently.

In DV quantum coherence resource theories defined by physically free operations \cite{chitambarpio}, or dephasing-covariant or translation-covariant quantum channels \cite{PhysRevA.94.052324}, the free operations can be characterized by their Stinespring dilations, i.e., by the interactions of the quantum system with an environment. This is not generically the case when free operations are given by the class of strictly incoherent (SIO) or incoherent operations (IO) \cite{chitambarpio,PhysRevA.94.052324,pleniocoherence,winterchitambar}. Because the class IO in a DV quantum coherence resource theory serves as the prototype for the free operations in QCRTF, it is therefore instructive to consider the question of whether the set $\mathcal{E}_{M}^{B}$ contains quantum channels defined via the following physical processes: 1. Appending a Fock state of auxiliary modes, 2. Applying a linear optical unitary on all modes, 3. Tracing out the auxiliary modes. Specifically, let $\mathcal{W}^{B}_{M}$ be the set of quantum channels $\eta: D(\mathcal{F}_{M})\rightarrow D(\mathcal{F}_{M})$ for which there is an $M'>M$, a set $E$ of $M'-M$ orthonormal modes, and an isometry $V:\mathcal{F}_{M}\rightarrow \mathcal{F}_{M'}$ such that
\begin{eqnarray}
{}&{}& \text{1. } \eta(\rho)= \text{tr}_{E}V\rho V^{*} \, , \, \nonumber \\
{}&{}& \text{2. }V\ket{\psi} \in \Delta_{M'}^{B} \,  \, \forall \, \ket{\psi}\in\Delta_{M}^{B}.
\label{eqn:stineset}
\end{eqnarray} 

In turns out that $\mathcal{W}^{B}_{M} \not\subset \mathcal{E}_{M}^{B}$, which implies that the physical processes discussed above do not have a simple relation to QCRTF$^{(B)}$. A counterexample to the inclusion is provided by a generalization of the Hong-Ou-Mandel effect that appears often in biphoton quantum optics experiments \cite{PhysRevLett.59.2044}. In particular, consider the symmetrized Fock spaces $\mathcal{F}_{2}$, $\mathcal{F}_{3}$ corresponding to the single particle Hilbert space with orthonormal basis $\lbrace \ket{1},\ket{2}\rbrace$ and $\lbrace \ket{1},\ket{2},\ket{3}\rbrace$ respectively. Also, consider an isometry $V:\mathcal{F}_{2}\rightarrow \mathcal{F}_{3}$ that acts in the one-photon sector as \begin{equation}Va^{*}_{\psi}\ket{\text{VAC}}= e^{i{\pi\over 4}(a_{1}^{*}a_{3}+h.c.)} a^{*}_{3}a^{*}_{\psi}\ket{\text{VAC}}\label{eqn:hom}\end{equation}
where $\ket{\psi}=c_{1}\ket{1}+c_{2}\ket{2}$, $\vert c_{1}\vert^{2} + \vert c_{2}\vert^{2}=1$.
Finally, construct the channel $\eta(\rho):= \text{tr}_{3}V\rho V^{*}$. Since one can formulate the set $\Delta^{B}_{3}$ starting from the orthonormal basis $\lbrace \ket{1},\ket{2},\ket{3}\rbrace$, it follows from (\ref{eqn:stineset}) with $M'=3$ and $E=\lbrace \ket{3}\rbrace$ that $\eta \in \mathcal{W}^{B}_{2}$. The channel $\eta$ satisfies 
\begin{align}\eta\left( a^{*}_{\psi}\ket{\text{VAC}}\bra{\text{VAC}} a_{\psi}\right) &= {\vert c_{1}\vert^{2}\over 2} \ket{\text{VAC}}\bra{\text{VAC}} \nonumber \\
&+ {\vert c_{2}\vert^{2}\over 2} \ket{0,1}\bra{0,1}  \nonumber \\
&+ {1\over 2}\ket{\phi}\bra{\phi}
\label{eqn:mandel}
\end{align}
where $\ket{\phi}:= \left( {c_{1}a_{1}^{*2}\over \sqrt{2}} + c_{2}a_{1}^{*}a_{2}^{*}\right) \ket{\text{VAC}}$ is a two-particle state. Note that if $c_{2}=0$, the common example of the Hong-Ou-Mandel effect is recovered.
Although the vacuum sector (respectively, one-particle sector) of the channel output is clearly in $Q_{0}\Delta_{M}^{B}Q_{0}$ (respectively, $Q_{1}\Delta_{M}^{B}Q_{1}$), the two-particle sector is in a pure state $\ket{\phi}$ that cannot be obtained by applying a number-conserving linear optical unitary to a Fock state on modes $\ket{1},\ket{2}$. Therefore, $\eta \notin \mathcal{E}^{B}_{2}$. The channel (\ref{eqn:hom}) is a specific example of a more general class of quantum channels in $\mathcal{W}^{B}_{M}$ called \textit{photon-added attenuators} \cite{PhysRevA.95.062309}, and form an important subset of Gaussian dilatable quantum channels. 

There are possibilities to amend the definitions of $\mathcal{E}_{M}^{B}$ and $\mathcal{W}^{B}_{M}$ so that the inclusion $\mathcal{W}^{B}_{M}\subset \mathcal{E}_{M}^{B}$ is obtained. We give an example, but do not discuss details because they are not relevant to the correlation distillation protocols in Section \ref{sec:ex}. In lieu of the third physical process (the partial trace) that led to the definition (\ref{eqn:stineset}) of $\mathcal{W}^{B}_{M}$, one can take the partial trace over $E$ defined as the orthonormal set of modes $\lbrace V\ket{M+1},\ldots,V\ket{M'}\rbrace$. In addition, in (\ref{eqn:embset}) one can consider the Kraus operators to map $\Delta_{M}^{B}$ to $\Delta_{M'}^{B}$ for $M'>M$. To see that these substitutions of definition give the desired inclusion, consider $\ket{\vec{n}}$ a Fock state on modes $\ket{M+1},\ldots , \ket{M'}$ and some pure state $U\ket{\vec{m}} \in \Delta_{M}^{B}$ (with $U$ a linear optical unitary on the modes $\ket{1},\ldots , \ket{M}$). Then, the state $VU\ket{\vec{m},\vec{n}} \in \Delta^{B}_{M'}$ is a Fock state on the orthonormal modes $VU\ket{1},\ldots , VU\ket{M},V\ket{M+1},\ldots , V\ket{M'}$. Therefore, with the present definition of $E$, $\text{tr}_{E}VU\ket{\vec{m},\vec{n}}\bra{\vec{m},\vec{n}}U^{*}V^{*} \in \Delta_{M'}^{B}$ because it is a Fock state on modes $VU\ket{1},\ldots , VU\ket{M}$ which span an $M$-dimensional subspace of $\text{span}_{\mathbb{C}}\lbrace \ket{1},\ldots,\ket{M'}\rbrace$.

\subsection{\label{sec:cc}QCRTF$^{(C)}$}

The final version of QCRTF that we consider aims to unify the finite-dimensional QCRT$^{(N)}_{M}$ and the quantum coherence resource theory of linear optics \cite{tanvolk}. The physical motivation for QCRTF$^{(C)}$ is provided by the consideration that quantum states defined by probability distributions over displacements of bosonic vacua are ``classical'', analogous to the definition of classical states in quantum optics as probability distributions over multimode Heisenberg-Weyl coherent states \cite{PhysRevLett.10.84,PhysRev.130.2529,PhysRev.131.2766,bach1986,perelomov}. The quantum coherence resource theory QCRTF$^{(C)}$ is defined by first considering the bosonic Fock space $\mathcal{F}_{M}$ with respect to $M$-dimensional Hilbert space $\mathcal{H}$, and the linear optical unitary channels $\mathcal{U}_{M}$. The set of free states $\Delta^{C}_{M}$ is then defined as follows: $\rho \in \Delta^{C}_{M}$ if any only if for any $0<\epsilon < 2$, there is an $N\in\mathbb{Z}_{\ge 0}$ and a joint probability density $p:  \bigoplus_{j=0}^{N}\left(\mathfrak{P}_{M}^{(j)} \times \mathbb{C}^{M} \times  U(M)\right) \rightarrow \mathbb{R}$  and $\rho'$ having the form
\begin{eqnarray}
\rho ' &=&  \bigoplus_{j=0}^{N}\sum_{\sigma^{(j)}} \int {d^{2}\vec{\alpha}_{j}\over \pi^{M}} \int \mu(dg_{j})\,p(\vec{\alpha}_{j},g_{j},\sigma^{(j)})U_{g_{j}}D(\vec{\alpha}_{j})\sigma^{(j)}D(\vec{\alpha}_{j})^{*} U_{g_{j}}^{*}
\label{eqn:freeC}
\end{eqnarray}
such that $\Vert \rho - \rho ' \Vert_{1} < \epsilon$.  In (\ref{eqn:freeC}), $\vec{\alpha}_{j}\in\mathbb{C}^{M}$, the phase space measure is defined as $d^{2}\vec{\alpha}_{j}/\pi^{M}=\prod_{\ell=0}^{M}{1\over \pi}d\text{Re}(\vec{\alpha}_{j})_{\ell}\, d\text{Im}(\vec{\alpha}_{j})_{\ell}$, and the displacement operator is defined by $D(\vec{\alpha}_{j}):=\exp iRZ_{\vec{\alpha}_{j}} $, where the column vector $Z_{\vec{\alpha}_{j}}\in \mathbb{R}^{2M}$ has components $z_{\ell}$ given by the solution of the linear equation $(\vec{\alpha}_{j})_{\ell}=(-z_{2\ell} + iz_{2\ell-1})/\sqrt{2}$, $\ell=1,\ldots,M$. Analogously to (\ref{eqn:freeB}), the sum over $\sigma^{(j)}$ in (\ref{eqn:freeC}) runs over a countable set of states in $\mathfrak{P}^{(j)}_{M}$ and the probability density is normalized as $\sum_{j=0}^{\infty}\sum_{\sigma^{(j)}}\int {d^{2}\vec{\alpha}_{j}\over \pi^{M}}\int \mu(dg_{j})p(\vec{\alpha}_{j},g_{j},\sigma^{(j)})=1$. However, unlike (\ref{eqn:freeB}), $\mu(dg_{j})$ is the normalized Haar measure on $U(M)$ for all $j\ge 0$. From the set of factorizable $p(\vec{\alpha}_{j} ,g_{j},\sigma^{(j)})$ with the form $p(\vec{\alpha}_{j} ,g_{j},\sigma^{(j)}) \propto \delta(\vec{\alpha}_{j})$ for all $j$ in (\ref{eqn:freeC}), it is clear that $\Delta^{B}_{M}$ is a proper subset of $\Delta^{C}_{M}$ (e.g., a displaced Fock state $D(\vec{\alpha})\ket{\vec{m}} \in \Delta^{C}_{M}$ but is not in $\Delta^{B}_{M}$ \footnote{Ref.\cite{nieto} details the argument that, like Heisenberg-Weyl coherent states, displaced Fock states retain the form of their wavefunction under time-evolution generated by $N_{M}$.}). By considering those $p(\vec{\alpha}_{j} ,g_{j},\sigma^{(j)})$ in (\ref{eqn:freeC}) that have support contained in the sector $j=0$,  one obtains the set of free states in the quantum coherence resource theory of linear optics (viz., $M$-mode classical states; those quantum states with $P$-function given by a probability density on $\mathbb{C}^{M}$).  However, the usual mathematical setting of quantum optics is a tensor product of a finite number of copies of $\ell^{2}(\mathbb{C})$, not the image of such a tensor product under the symmetrizer $P_{B}$. The connection between $\Delta_{M}^{C}$ and quantum optics is made by going from the symmetrized Fock representation $\mathcal{F}_{M}$ to the Schr\"{o}dinger representation $L^{2}(\mathbb{R}^{M})$ of the canonical commutation relations \cite{robinsonvol2}, i.e., by mapping the bosonic state $\ket{\vec{m}} \in \Delta^{(N)}_{M}$ to $\ket{m_{1}}\otimes \cdots \otimes \ket{m_{M}}$ of an $M$-dimensional quantum harmonic oscillator. Finally, within a given particle number sector, a pure state in $\Delta^{C}_{M}$  may be a superposition of states of the form $U\ket{\vec{m}}$ with $U\in \tilde{\mathcal{U}}_{M}$. This is a significant difference from $\Delta^{(A,B)}_{M}$. However, the amplitudes of the superposition must be derivable from statistics that are consistent with application of a displacement operator $D(\vec{\alpha})$ to a pure state in $\Delta_{M}^{B}$ \cite{nieto}. This condition is analogous to the verification of non-classicality in quantum optics by verification of non-Poissonian statistics of a photon counting measurement. As an example of this possibility, one can consider
\begin{align}
Q_{2}e^{\alpha a_{1}^{*}-\overline{\alpha}a_{1}}e^{\alpha a_{2}^{*}-\overline{\alpha}a_{2}}\ket{\text{VAC}}&\propto a_{1}^{*}a_{2}^{*}\ket{\text{VAC}}-iUa_{1}^{*}a_{2}^{*}\ket{\text{VAC}}
\end{align}
with $U=e^{i{\pi \over 4}(a_{1}^{*}a_{2}+h.c.)}$.

A set of free operations $\mathcal{E}_{M}^{C}$ in QCRTF$^{(C)}$ is constructed by a method analogous to the set $\mathcal{E}_{M}^{B}$ in QCRTF$^{(B)}$. Specifically, the set $\mathcal{E}_{M}^{C}$ is defined by replacing ``B'' with ``C'' in (\ref{eqn:embset}). A subset of physically free quantum channels in $\mathcal{E}^{C}_{M}$ can be constructed analogously to the modified version of $\mathcal{W}^{B}_{M}$ discussed in the last paragraph of Section \ref{sec:bb}. Because we do not explore information processing tasks carried out with operations from $\mathcal{W}^{(B,C)}_{M}$, we consider the details of their structure theory a topic for future research.

\subsection{\label{sec:physconsid}Coherence quantifiers and physical considerations}

We now describe methods for quantifying coherence in QCRTF$^{(A,B,C)}$, focusing on QCRTF$^{(A)}$ because we will utilize free operations from $\mathcal{E}^{A}_{M}$ in the state distillation protocols in Section \ref{sec:ex}. Quantum states in $D(\mathcal{F}_{M})$ exhibit coherence with respect to QCRTF$^{(A)}$ in two ways, viz., a state can have coherence between particle number sectors or within a particle number sector. For example, in QCRTF$^{(A)}$, average coherence within the particle number sectors can be defined via the weighted entropy of coherence $C(\rho)=\sum_{N\ge 0}(\text{tr}Q_{N}\rho ) C_{N}(\rho)$ where $C_{N}(\rho)=H(\sum_{\sigma\in \mathfrak{P}^{(N)}_{M}}\sigma {Q_{N}\rho Q_{N} \over \text{tr}Q_{N}\rho} \sigma) - H({Q_{N}\rho Q_{N} \over \text{tr}Q_{N}\rho} )$, where $H$ is the von Neumann entropy function on quantum states \cite{winteryang}. On the other hand, the total coherence can be quantified by $C^{A}(\rho)=\lim_{N\rightarrow \infty}H(\sum_{\sigma\in \mathfrak{P}^{(0)}_{M}\oplus \cdots \oplus \mathfrak{P}^{(N)}_{M} } \sigma \rho \sigma ) - H(\rho)$. Quantum states in $D(\mathcal{F}_{M})$ can exhibit coherence with respect to QCRTF$^{(B,C)}$ if they exhibit superposition in the basis of coherent states with various bosonic vacua, for example, spin-squeezed states in the case of QCRTF$^{(B)}$ and quadrature squeezed states in the case of QCRTF$^{(C)}$. For QCRTF$^{(B,C)}$, coherence measures that satisfy widely adopted axioms for the quantification of quantum coherence can be constructed by generalizing the algorithm in Ref.\cite{tanvolk}. Since our examples in Section \ref{sec:ex} make use of the aforementioned entropy of coherence $C$ and $C^{A}$, we do not flesh out the details of extending the algorithm here.

We conclude this subsection with a discussion of some features of the QCRTF$^{(A,B,C)}$ that are relevant to the physical aspects of information processing within these theories. Firstly, unlike for finite-dimensional quantum coherence resource theories on spin-$1/2$ chains, the CNOT operation acting on two particles is not free in any of the QCRTF because it is not a linear transformation of the bosonic Fock space (e.g., it maps the bosonic state $a_{0}^{*}a_{1}^{*}\ket{\text{VAC}} = {1\over \sqrt{2}}(\ket{0}_{A}\otimes \ket{1}_{B} + \ket{1}_{A}\otimes \ket{0}_{B})$ to ${1\over\sqrt{2}}( \ket{0}_{A}\otimes \ket{1}_{B} + \ket{1}_{A}\otimes \ket{1}_{B})$). The incoherent nature of the CNOT operation is central to analyses of the relation between coherence and entanglement in DV systems \cite{PhysRevA.98.052351,PhysRevA.97.052304}. 
Finally, a principal criticism regarding the definition of free operations as given in (\ref{eqn:freeopA}) and (\ref{eqn:embset}) is that there is no \textit{a priori} constraint on the energy density of the set of free states. For example, in QCRTF$^{(A)}$ there are free quantum channels that transform the homogeneous state $\ket{1,1,\ldots, 1}$ to $\ket{M,0,\ldots ,0}$, whereas such an operation is not always feasible in a given optical trap setup. In the protocols discussed in Section \ref{sec:ex} for distillation of maximally correlated states from initial states in $D(\mathcal{F}_{2})$, we only consider free quantum channels that preserve the energy density, i.e., preserve the expected local number of particles with respect to a specified partition of the single particle modes (even if the total number of modes varies).

\section{Maximally correlated state distillation\label{sec:ex}}

We now consider protocols for distillation of maximally correlated states within the QCRTF framework which involve state preparation protocols in $\mathcal{E}^{B}_{M}$ and measurement protocols in $\mathcal{E}^{A}_{M}$ (Section \ref{sec:nc}) or preparation of states of maximal coherence (as quantified by $C^{A}$) followed by measurement protocols in $\mathcal{E}^{A}_{M}$ (Section \ref{sec:nonnumcons}). The two-mode maximally correlated states (\ref{eqn:mcmc}) and (\ref{eqn:mctilde}) which are the distillation targets maximize both the quantum mutual information between the two modes \cite{wilde} and mode entanglement \cite{PhysRevLett.91.097902} within each particle number sector. The measurements in $\mathcal{E}^{A}_{M}$ that we utilize are local to pairs of modes and, further, preserve the local particle number in expectation, in accordance with the discussion in Section \ref{sec:physconsid}. The final step of the distillation protocols involves a free, non-local isometry which maps Fock states on $2n$ modes to Fock states on $2nR$ modes ($R$ is the rate). This step contrasts with the isometry implemented in entanglement concentration, which is required to be local because the free operations in the resource theory of entanglement are LOCC \cite{nielsen00,everythingyou}.

Other quantum information processing protocols that can be analyzed using the QCRTF framework include the distillation of Gaussian entanglement from non-Gaussian states reported in Ref.\cite{pleniobrowne}, which involves only linear optical unitary operations (which are in $\mathcal{E}_{M}^{(B,C)}$) and local vacuum projections (which are in $\mathcal{E}_{M}^{A}$; vacuum projections on $M-1$ modes are in $\mathcal{E}^{(A,B,C)}_{M}$). Other protocols that implement unitary dynamics in $\mathcal{E}_{M}^{(B,C)}$ supplemented by particle detection measurements in $\mathcal{E}_{M}^{A}$ include boson sampling \cite{aaronson}, and cat state amplification \cite{PhysRevA.70.020101}. 

\subsection{\label{sec:nc}Number-conserving correlation distillation}

We start with the task of producing maximal correlations from bosonic states of fixed particle number $N$ within the QCRTF framework. Specifically, starting from $n$ bosonic copies of a state in $Q_{N}\Delta_{2}^{B}Q_{N}$ we consider using quantum channels from $\mathcal{E}_{2n}^{A}$ to produce $\lfloor nR \rfloor$ bosonic copies of a maximally correlated bosonic state of two modes and $N$ particles. The target state of the distillation protocol consists of $\lfloor nR \rfloor$ bosonic copies of a two-mode maximally correlated state which we take as \begin{equation} \ket{\text{MC}_{N}}=(N+1)^{-1/2} \sum_{m=0}^{N}\ket{N-m,m} \label{eqn:mcmc} \end{equation} without loss of generality. By $\tilde{n}:= \lfloor nR \rfloor$ \textit{bosonic copies} of $\ket{\text{MC}_{N}}$, we mean the $2\tilde{n}$ mode state
\begin{equation}
\ket{\text{MC}_{N,n}}\propto \sum_{m_{1}=0}^{N}\cdots\sum_{m_{n}=0}^{N}\ket{N-m_{1},m_{1},N-m_{2},m_{2},\cdots,N-m_{\tilde{n}},m_{\tilde{n}}}
\label{eqn:mclab}
\end{equation} and not the non-bosonic state $\ket{\text{MC}_{N}}^{\otimes \tilde{n}}$. To make the difference between $\ket{\text{MC}_{N}}^{\otimes \tilde{n}}$ and $\ket{\text{MC}_{N,\tilde{n}}}$ clear, consider writing $\ket{\text{MC}_{N}}^{\otimes 2}$ in terms of $2N$ qubits as \begin{equation}\ket{\text{MC}_{N}}^{\otimes 2} \propto \left( P_{B}\sum_{k=0}^{N}\ket{1}^{\otimes N-k}\otimes \ket{2}^{\otimes k}\right)\otimes \left( P_{B}\sum_{k=0}^{N}\ket{1}^{\otimes N-k}\otimes \ket{2}^{\otimes k}\right).\label{eqn:mcfake} \end{equation} The state (\ref{eqn:mcfake}) is comprised of only 2 single-particle modes (viz., $\ket{1}$ and $\ket{2}$), and, further, is not bosonic because the full $2N$-qubit state has not been symmetrized. Unlike $\ket{\text{MC}_{N}}^{\otimes \tilde{n}}$, the state $\ket{\text{MC}_{N,\tilde{n}}}$ in (\ref{eqn:mclab}) that we desire to distill is a fully bosonic state with $2\tilde{n}$ modes $\lbrace \ket{1},\ldots ,\ket{2\tilde{n}} \rbrace$ such that the maximal correlations between mode pair $(\ket{1},\ket{2})$ are the same as for $(\ket{3},\ket{4})$, etc., as could occur in an ultracold optical lattice with $2\tilde{n}$ sites. Note that  $\text{MC}_{1} \in \Delta_{2}^{(B,C)}$ since it is simply a maximally coherent qubit state, but for $N>1$, $\text{MC}_{N} \notin \Delta_{2}^{(B,C)}$.  By imposing the constraint that initial states and target state have the same particle number, the distillation rate provides an asymptotically energy-independent quantity that characterizes the usefulness of the initial state as a resource for producing quantum correlations.

\begin{figure}[t]
\centering
\includegraphics[scale=0.4]{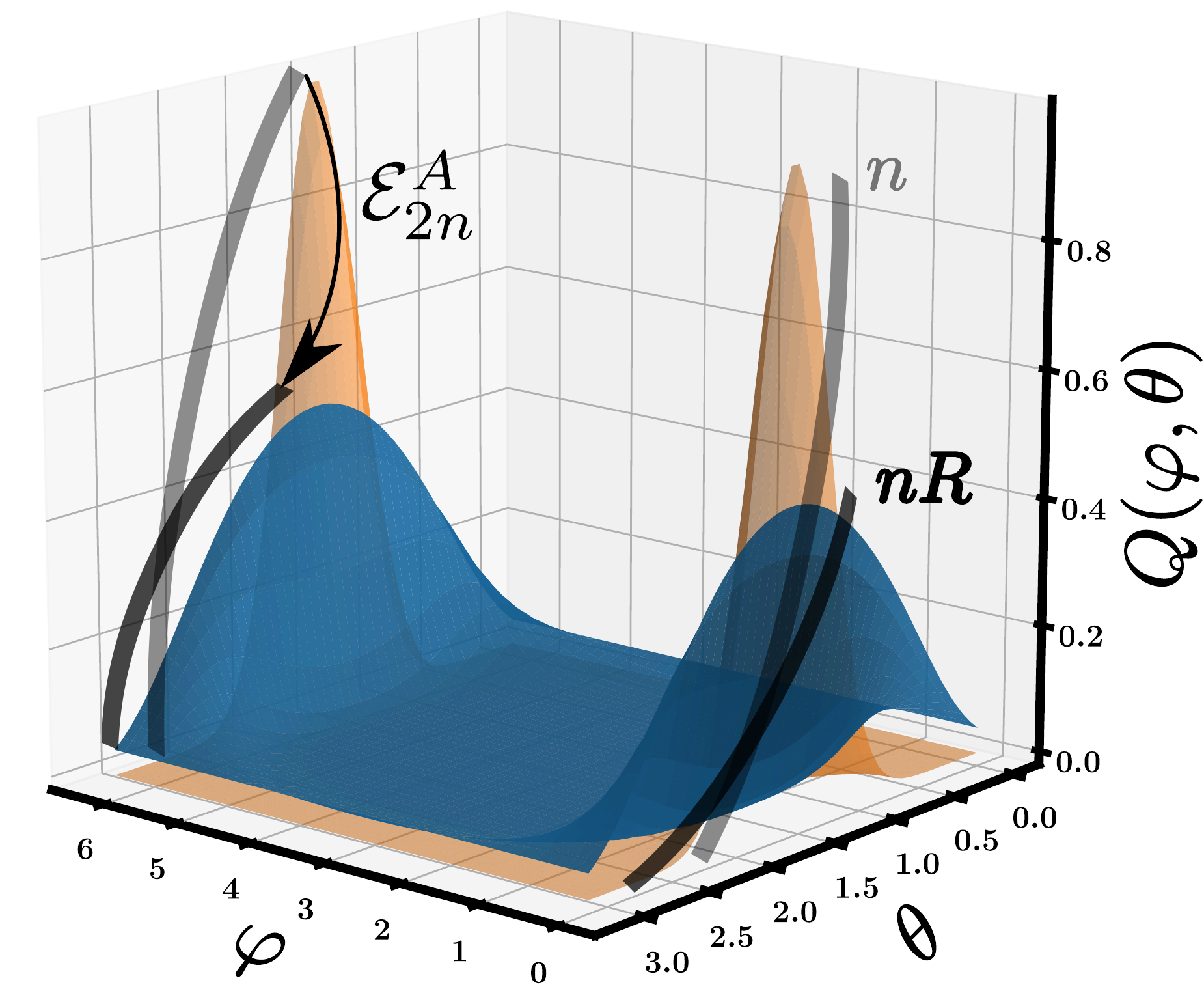}
\caption{Schematic of number-conserving distillation protocol of $nR$ bosonic copies of $\ket{\text{MC}_{N,1}}$ (blue surface) from $n$ bosonic copies of $\ket{\text{BEC}_{N,1}}$ (orange surface). $N=24$ is used in the figure. The surfaces are the respective $SU(2)$ $Q$-functions of the states, which can also be visualized on the 2-sphere \cite{PhysRevA.47.5138}.
}
\label{fig:qu} 
\end{figure}

To begin, consider an input state comprised of $n$ bosonic copies of a Bose-Einstein condensed state describing $N$ particles condensed into an equal-amplitude superposition $(1/\sqrt{2})(\ket{1}+\ket{2})$ of two orthogonal single-particle states $\ket{1}$ and $\ket{2}$. The process of taking $n$ bosonic copies extends the orthonormal basis to $\lbrace \ket{1},\ldots ,\ket{2n}\rbrace$ so the described input state can be written\begin{equation}\ket{\text{BEC}_{N,n}}\propto \prod_{j=1}^{n}(a_{2j-1}^{*}+a_{2j}^{*})^{N}\ket{\text{VAC}} \in \Delta_{2n}^{B}.  \label{eqn:bec}\end{equation}
This is a bosonic version of $n$ \textit{independent, identically distributed} Bose-Einstein condensates. Specifically, in a traditional quantum communication task, $n$ independent copies of the Bose-Einstein condensate $\left( \ket{1}+\ket{2}\over \sqrt{2}\right)^{\otimes N}$ would give the $nN$-partite tensor product state $\left( \ket{1}+\ket{2}\over \sqrt{2}\right)^{\otimes nN}$, which is not a suitable notion of $n$ independent copies of a Bose-Einstein condensate of $N$ particles because the $n$ copies are not individually addressable. It is preferable to prepare $n$ copies by spatially copying the Bose-Einstein condensate into $n$ pairs of modes (in the context of optically trapped ultracold bosons, increasing the number of optical lattice sites from $2$ to $2n$). Therefore, one may consider \begin{equation}\left( {\ket{1}+\ket{2}\over \sqrt{2}} \right)^{\otimes N}\otimes \left( {\ket{3} +\ket{4} \over \sqrt{2}} \right)^{\otimes N} \otimes \cdots \otimes \left( {\ket{2n-1}+\ket{2n} \over \sqrt{2}}\right)^{\otimes N}\label{eqn:gggg}\end{equation} to suitably describe $n$ independent copies of an $N$ particle Bose-Einstein condensate. But (\ref{eqn:gggg}) is not a bosonic state, since it is not a symmetric state of $nN$ particles. Upon projecting (\ref{eqn:gggg}) to $\mathcal{F}_{2n}^{(Nn)}$, one obtains $\ket{\text{BEC}_{N,n}}$ in (\ref{eqn:bec}). Fig. (\ref{fig:qu}) shows the $SU(2)$ $Q$-functions \cite{PhysRevA.47.5138} for $\ket{\text{BEC}_{N,1}}$ and $\ket{\text{MC}_{N,1}}$ and schematically shows the distillation protocol.

In order to derive the optimal asymptotic rate $R$ for distillation of an $\text{MC}_{N,\tilde{n}}$ state from $\text{BEC}_{N,n}$ as $n\rightarrow \infty$, we proceed by analogy with the optimal protocol for pure state distillation in the quantum coherence resource theory of spin chains \cite{PhysRevA.92.022124,winteryang}. We denote by $\vec{m}^{n}$ a list $(\vec{m}_{1},\ldots ,\vec{m}_{n})$ such that $\vec{m}_{j} = (m_{j,1},m_{j,2}) \in\mathbb{Z}_{\ge 0}^{\times 2}$, and $\vert \vec{m}_{j} \vert = N$. It is clear that every projection onto a type class \cite{wilde} $T_{t}:=\lbrace \vec{m}^{n} : t_{\vec{m}^{n}}=t\rbrace$ defines a state in $\Delta_{2n}^{A}$ (specifically, in $Q_{nN}\Delta_{2n}^{A}Q_{nN}$), and therefore the type class measurement $\lbrace \sum_{\vec{m}^{n}\in T_{t}}\ket{\vec{m}^{n}}\bra{\vec{m}^{n}} \rbrace_{t}$ is in $\mathcal{E}^{A}_{2n}$.  Because the amplitudes of the state $\text{BEC}_{N,1}$ in the Fock basis are the square root of the binomial distribution $B(N,1/2)$, the probability of obtaining the result $\vec{m}_{j}$ upon measurement of $\ket{\text{BEC}_{N,n}}$ in the Fock basis on modes $(\ket{2j-1},\ket{2j})$ is given by the binomial distribution $B(N,1/2)$, for every $j=1,\ldots,n$. It follows from the law of large numbers that for arbitrarily small $\delta >0$, the type class measurement will produce for $n\rightarrow \infty$ a uniform superposition of $\ket{\vec{m}^{n}}$ with $\vec{m}^{n}$ a $\delta$-typical sequence. For a probability distribution $p$, we refer to the set of $\delta$-typical sequences as $\delta \rightarrow 0$ as $\mathcal{T}_{p}$. Therefore, with $p=B(N,1/2)$ and $n\rightarrow \infty$, application of the typical subspace measurement to the state $\text{BEC}_{N,n}$ gives the state
\begin{equation}
{1\over \sqrt{2^{nH(B(N,1/2))}}}\sum_{\vec{m}^{n}\in \mathcal{T}_{B(N,1/2)}\atop \vert \vec{m}_{j} \vert = N}\ket{\vec{m}^{n} }
\label{eqn:bectype}
\end{equation}
with probability going to $1$, where for a discrete probability density $p$, $H(p)$ is the Shannon entropy. Note that $\ket{\vec{m}^{n}}\in \Delta^{A}_{2n}$. To convert the state (\ref{eqn:bectype}) to $nR$ copies of a $\text{MC}_{N}$ state, one defines an isometry that takes $\ket{\vec{m}^{n}}$ to a state of the form $\ket{\vec{\ell}^{\,\log_{N+1}\vert \mathcal{T}_{B(N,1/2)}\vert}} \in \Delta^{A}_{2\log_{N+1}\vert T_{B(N,1/2)}\vert}$, where $\vec{\ell}_{j} = (\ell_{j,1},\ell_{j,2}) \in \lbrace (N-r,r) \rbrace_{r=0}^{N}$ for $j=1,2,\ldots,\log_{N+1}\vert \mathcal{T}_{B(N,1/2)}\vert$. Such an isometry maps the state in (\ref{eqn:bectype}) to the state $\text{MC}_{N,nR}$ in (\ref{eqn:mclab}) with $nR=\log_{N+1}\vert \mathcal{T}_{B(N,1/2)}\vert$. Both the typical subspace measurement and the isometry preserve the expected particle number $N$ in each mode pair $(\ket{2j-1},\ket{2j})$, therefore keeping the energy density constant. The optimal rate $R$ is given by 
\begin{align}R &={1\over n}\log_{N+1}\vert \mathcal{T}_{B(N,1/2)}\vert  \nonumber \\
&={H(B(N,1/2)) \over  \log_{2}(N+1) }\nonumber \\
&\rightarrow {1\over 2} \; \text{as} \; N\rightarrow \infty  
\end{align}
where, in practice, $\tilde{n}=\lfloor nR \rfloor$ modes occur in the distilled state. Note that if the two orthogonal single particle modes that define  $\text{MC}_{N}$ are are taken to be different from the initial modes of $\text{BEC}_{N}$, the required unitary rotation can be freely implemented in $\mathcal{E}^{B}_{2n}$. Finally, it follows from the asymptotics of the Shannon entropy of the multinomial distribution \cite{cichon} that replacing the two mode, $N$ particle BEC in (\ref{eqn:bec}) by a $M$ mode, $N$ particle BEC gives an optimal rate of $\text{MC}_{N}$ distillation that scales linearly with $M$.

\begin{figure}[t]
\centering
\includegraphics[scale=0.8]{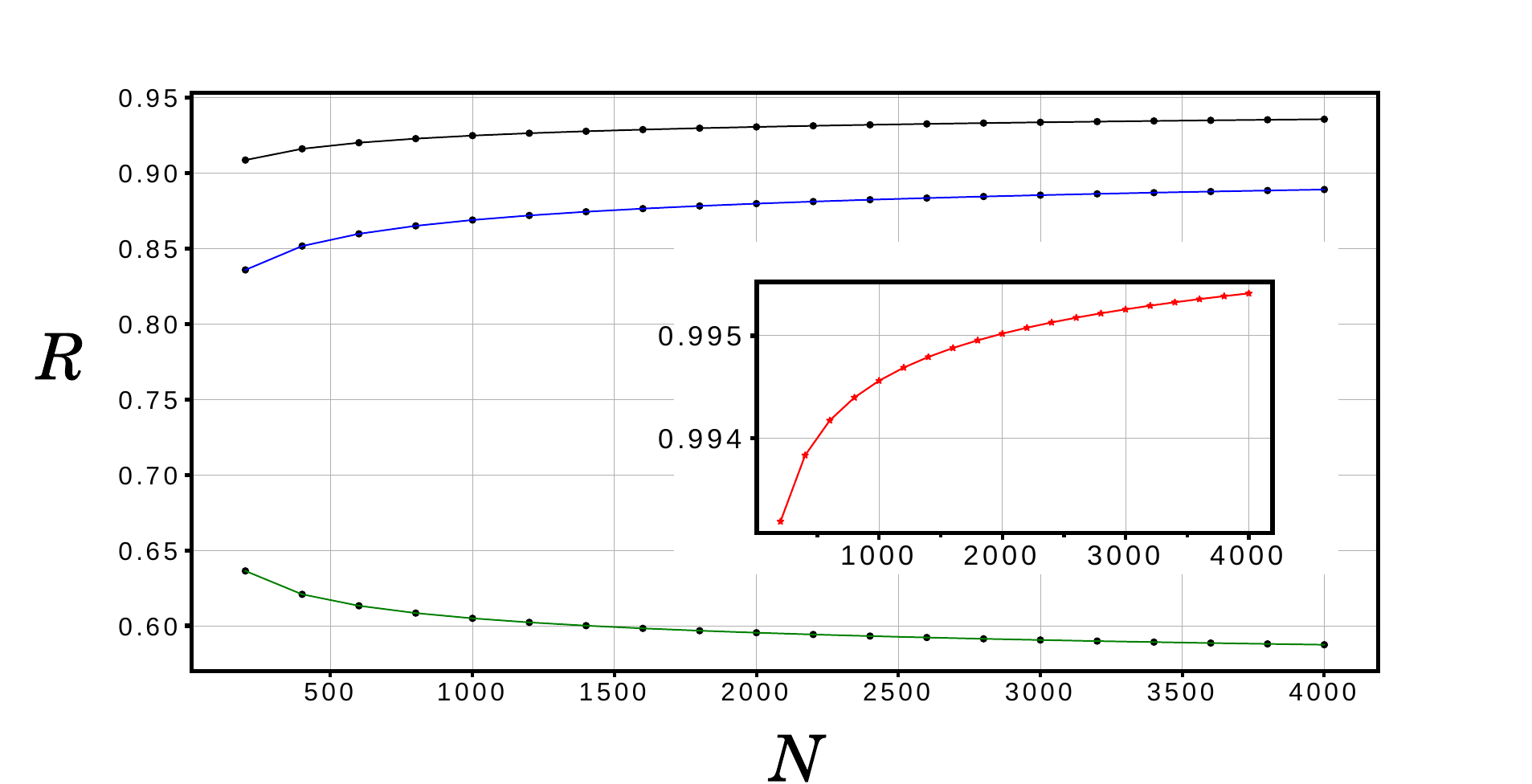}
\caption{Optimal distillation rate of maximally correlated state MC$_{N}$ from $\Psi(\theta= \pi/4,m)_{N}$ ($N$ up to 4000). The data shown correspond to $m=0$ (green), $m=N/2$ (blue), and $m=N/2-1$ (black). Inset: optimal distillation rate of $\widetilde{\text{MC}}_{N}$ from maximally coherent state $\Phi_{N}$ with respect to the same abscissa as the main figure.
}
\label{fig:fits} 
\end{figure}

This example shows that Bose-Einstein condensates are poor resources for extraction of intermode correlation within QCRTF, and leads to natural questions concerning which states allow an optimal asymptotic rate of maximal correlation production. For instance, one may seek elements of $Q_{N}\Delta^{B}_{2}Q_{N}$ that maximize the $\text{MC}_{N}$ distillation rate, or at least serve as better resources than BEC$_{N,1}$. 

With this motivation, we consider the optimal asymptotic distillation rate of $\text{MC}_{N}$ over all $Q_{N}\Delta^{B}_{2}Q_{N}$. As seen from the previous example, the distillation rates for states of fixed particle number are equal to the entropy of coherence $C$ with respect to QCRTF$^{(A)}$ (and scaled by an appropriate factor) \cite{winteryang,PhysRevA.92.022124}. The quantity $C$ is convex on the quantum state space and invariant under argument shifts of complex Fock state amplitudes. Therefore, the search can be restricted to pure states in $Q_{N}\Delta^{B}_{2}Q_{N}$ with real amplitudes, without loss of generality. These states take the form $\Psi(\theta ,m)_{N} \propto \left( a_{1}^{*}+\tan \theta a_{2}^{*} \right)^{m}\left( a_{1}^{*}-\tan \theta a_{2}^{*} \right)^{N-m}\ket{\text{VAC}}$, and we search over $\theta \in [0, \pi/4]$ and $m\in \lbrace 0,1,\ldots ,N/2 \rbrace$ due to the symmetry of the factors. From numerical computation, we find that the greatest entropy of coherence over $Q_{N}\Delta^{B}_{2}Q_{N}$ occurs for $\theta = \pi / 4$, but neither for $m=0$ (viz., $\text{BEC}_{N,1}$) nor $m=N/2$ (viz., the pair correlated state $\left( a_{1}^{*2}+ a_{2}^{*2} \right)^{{N\over 2}}  \ket{\text{VAC}}$ \cite{PhysRevA.96.023621}). In Fig.\ref{fig:fits} we show the entropy of coherence data for $N$ up to 4000 for $\theta = \pi / 4$ and selected values of $m$. Although $m=N/2$, $\theta = \pi / 4$ does not give the greatest entropy of coherence over $Q_{N}\Delta^{B}_{2}Q_{N}$ (and, therefore, does not allow the greatest $\text{MC}_{N}$ distillation rate in $Q_{N}\Delta^{B}_{2}Q_{N}$), it is an interesting case for two reasons: 1. the $C(\Psi(\pi/4 , N/2)_{N}) \sim \log_{2}N^{x}$ scaling  with $x>1/2$ can be analytically verified (see Appendix \ref{sec:pcsanal}), and 2. this state can be obtained from a two-mode CV quadrature squeezed state by applying a beamsplitter followed by the particle number sector projection $Q_{N}$, which indicates that maximal intermode correlations can be distilled at a greater rate from quadrature squeezed states than from Heisenberg-Weyl coherent states using operations from $\mathcal{E}^{A}_{2}$.

\subsection{\label{sec:nonnumcons}Non-number-conserving correlation distillation}

We now consider utilizing $n$ bosonic copies of a state of indefinite particle number to distill $\lfloor nR \rfloor$ bosonic copies of the following maximally correlated state of indefinite particle number
\begin{equation}
\ket{\widetilde{\text{MC}}_{N}}:={1\over \sqrt{2N+1}}\sum_{k=0}^{2N}\ket{\text{MC}_{k}} \in \mathcal{F}_{2}.
\label{eqn:mctilde}
\end{equation}
with asymptotically optimal rate $R$, again using only quantum channels from $\mathcal{E}^{A}_{2n}$.
Using the notation $\widetilde{\text{MC}}_{N} := \ket{\widetilde{\text{MC}}_{N}}\bra{\widetilde{\text{MC}}_{N}}$, one finds that $\mathcal{N}_{2}(\widetilde{\text{MC}}_{N}) = N$. Although it is possible to calculate the distillation rate of $\ket{\widetilde{\text{MC}}_{N}}$ starting from a CV coherent state in $\Delta^{C}_{2}$, which is analogous to the case of $\text{BEC}_{N,1}$ considered in Subsection \ref{sec:nc}, the distillation rate is not expected to be optimal because the value of the coherence measure $C^{A}$ in such states (see Section \ref{sec:physconsid}) is not maximal. Instead, we focus in this section on distillation of $\lfloor nR \rfloor$ bosonic copies of $\ket{\widetilde{\text{MC}}_{N} }$ from $n$ bosonic copies of a resource state which exhibits maximal coherence with respect to QCRTF$^{(A)}$, i.e., a state $\rho\in D(\mathcal{F}_{2})$ that exhibits the maximum value of coherence $C^{A}$ subject to the expected particle number constraint $\mathcal{N}_{2}(\rho)=N$. To identify a state with maximal coherence subject to the expected particle number constraint, we utilize the Karush-Kuhn-Tucker conditions \cite{cover} to prove Proposition \ref{prop:prop3} (the proof is given in Appendix \ref{sec:kkt}).

\begin{proposition}\label{prop:prop3} A state $\rho \in D(\mathcal{F}_{2})$  that achieves the maximum value of $C^{A}(\rho)$ subject to the constraint $\mathcal{N}_{2}(\rho)=N$ is given by
\begin{equation}
\ket{\Phi_{N}}:=\sum_{k=0}^{\infty}{2\over N+2} \left( {N\over N+2 }\right)^{k\over 2}\sum_{r=0}^{k}\ket{k-r,r} .
\label{eqn:maxeoc}
\end{equation}
\end{proposition}

Note that the amplitudes of $\Phi_{N}$ are not uniform across the particle number sectors. Further, it is clear that $Q_{N}\Phi_{N}Q_{N}\propto \text{MC}_{N}$, so $\Phi_{N}$ is maximally correlated in each particle number sector. In Appendix \ref{sec:entmaxanal}, we derive the asymptotic scaling $C^{A}(\Phi_{N}) \sim \log_{2}N^{2}$, which shows that superposition of particle number sectors allows a greater maximal coherence compared to the case of fixed particle number analyzed in the previous section. There, we also derive $C^{A}(\widetilde{\text{MC}}_{N})$ for completeness.

A general state $\ket{G_{N}} \in \mathcal{F}_{2}$ such that $\mathcal{N}_{2}(G_{N})=N$ defines a joint probability distribution $p_{X,Y}(x,y)$ on the subset of $\mathbb{Z}^{\times 2}$ defined by $\lbrace (x,y): \, x\ge 0 \, , \, 0\le y \le x \rbrace$. This can be seen by writing $\ket{G_{N}}=\sum_{x=0}^{\infty}\sum_{y=0}^{x}\sqrt{p_{X,Y}(x,y)} \ket{x-y,y}$. Note that $X$ and $Y$ are not necessarily independent random variables. Starting from $n$ bosonic copies of $\ket{G_{N}}$ (formed in the same way as in (\ref{eqn:bec}), (\ref{eqn:mclab})) one applies the $\delta$-typical measurement (in $\mathcal{E}_{2n}^{A}$) defined by the set of jointly $\delta$-typical sequences $T^{\delta}$ corresponding to $p_{X,Y}(x,y)$, which produces the following state of indefinite particle number with high probability for large $n$:
\begin{equation}
{1\over \sqrt{ \vert  T^{\delta} \vert}}\sum_{\vec{n}\in T^{\delta}}\ket{\vec{n}^{n}}
\label{eqn:gentyp}
\end{equation}
where $\vec{n} \in \lbrace (N' -r , r): \, N'\in \mathbb{Z}_{\ge 0} \, , \, r \in \lbrace 0,\ldots ,N' \rbrace \rbrace$.  The final step is to apply an isometry that maps $\ket{\vec{n}^{n}}$ (with $\vec{n}^{n}\in T^{\delta}$) to \begin{equation} \ket{\vec{\ell}^{\log_{(2N+1)(N+1)}\vert T^{\delta} \vert}} \in \Delta_{2\log_{(2N+1)(N+1)}\vert T^{\delta} \vert}^{A},\end{equation} where for any $n'$ we have defined $\vec{\ell}^{n'}:= (\vec{\ell}_{1},\ldots ,\vec{\ell}_{n'})$ such that $\vec{\ell}_{j}  \in\mathbb{Z}_{\ge 0}^{\times 2}$, and $\vert \vec{\ell}_{j} \vert \in \lbrace 0,\ldots , 2N\rbrace$ due to the upper limit $2N$ on the particle number sector of MC$_{N}$. Note that the image of this isometry has dimension $\vert T^{\delta} \vert$ and that the expected number of particles in each mode pair is $N$, which shows that the isometry also conserves the energy density. The output of the isometry is $\log_{(2N+1)(N+1)}\vert T^{\delta} \vert$ bosonic copies of the state $\ket{\widetilde{\text{MC}}_{N}}$. With $\delta \rightarrow 0$, it follows that the distillation rate for $n\rightarrow \infty$ is \begin{equation}
R= (\log_{2}(2N+1)(N+1))^{-1}H(p_{X,Y}).
\label{eqn:rmaxcoh}
\end{equation}
For $\ket{G_{N}}$ given by $\ket{\Phi_{N}}$ in (\ref{eqn:maxeoc}), this rate is shown in the inset of Fig.\ref{fig:fits}. The result shows that the state with maximal coherence with respect to QCRTF$^{(A)}$ (i.e., maximal $C^{A}$)  can be used to extract a uniform superposition of maximally correlated states with unit efficiency as $n\rightarrow \infty$.

\section{Conclusion}In this work, we have introduced and analyzed three versions of QCRTF that allow a wide variety of DV and CV quantum information processing protocols to be considered within a unified resource theoretic framework. QCRTF$^{(B,C)}$ are of particular interest due to the fact that the respective sets of free quantum states do not change after applying linear optical unitaries. The present results on optimal distillation of maximally correlated states indicate that quantum coherence with respect to QCRTF$^{(A)}$ is a resource for producing such correlations. Because the sets of free quantum states and free operations in the QCRTF framework are not limited to traditional classical states (e.g., states with positive $P$ function) and classical dynamics (e.g., Fokker-Planck dynamics of the $P$ function), respectively, they may allow to circumvent certain no-go theorems of other many-body resource theories that utilize such a classical or quasiclassical framework, such as the impossibility of Gaussian entanglement distillation by Gaussian local operations supplemented by classical communication \cite{PhysRevA.66.032316}. We note that it is also possible to break the constraint of energy-density-preserving free operations in Section \ref{sec:ex} or consider asymptotic distillation rates of other entangled states, e.g., GHZ states of $N$ particles in two orthogonal modes, within the QCRTF framework. In such cases, the distillation rate may exhibit asymptotically non-constant scaling with $N$. Other potential directions of future research include analyses of fully bosonic one-shot distillation and formation \cite{winterchitambar}, and localization of the QCRTF \cite{PhysRevB.98.014519,PhysRevA.100.022331}.  The results reported here are expected to stimulate future work on optimal distillation of many-body DV and CV quantum states that are useful for quantum communication, quantum error correction, or quantum metrology \cite{yadin,PhysRevLett.122.040503,PhysRevA.94.042327,volkphotonic}, within the framework of experimentally-motivated QCRTs.

\section*{Acknowledgments}

This work was supported by the National Research Foundation of Korea (NRF) funded by the
Ministry of Science and ICT (Grant No. 2016H1D3A1908876) and from the Basic Science Research Program through the NRF funded by the Ministry of Education (Grant No. 2015R1D1A1A09056745). The author also acknowledges support from the LDRD program at LANL. Los Alamos National Laboratory is managed by Triad National Security, LLC, for the National Nuclear Security Administration of the U.S. Department of Energy under Contract No. 89233218CNA000001.

%
\nocite{apsrev41Control}
\bibliography{phasebib2,revtex-custom}

\onecolumn\newpage
\appendix

\section{\label{sec:kkt}Proof of Proposition \ref{prop:prop3}}

Consider maximization of $C^{A}$ for a pure quantum state of the form \begin{equation}\ket{\Phi}=\sum_{k=0}^{\infty}\sqrt{p_{k}}\sum_{m=0}^{k}{1\over \sqrt{k+1}}\ket{k-m,m} \in \mathcal{F}_{2}\end{equation} subject to $\sum_{k=0}^{\infty}p_{k}=1$ (normalization), and $\sum_{k=0}^{\infty}kp_{k}=N$ (energy constraint $\mathcal{N}_{2}(\ket{\Phi})=N)$). The extremal condition is
\begin{equation}
\del_{p_{j}}\left( C^{A}(\ket{\Phi})+\lambda_{0}\sum_{k=0}^{\infty}p_{k} + \lambda_{1}\sum_{k=0}^{\infty}kp_{k} \right)=0
\end{equation}
and we evaluate (using natural logarithm for convenience)
\begin{align}
C^{A}(\ket{\Phi})&= -p_{0}\ln p_{0}-\sum_{k=1}^{\infty}\sum_{m=0}^{k}p_{k}(k+1)^{-1}\ln \left(p_{k}(k+1)^{-1}\right) \nonumber \\
&=-p_{0}\ln p_{0}-\sum_{k=1}^{\infty}p_{k}\ln p_{k}+\sum_{k=1}^{\infty}p_{k}\ln \left( k+1 \right).
\end{align}

The extremal condition is solved by $p_{0}=p_{0}^{(*)}=e^{\lambda_{0}-1}$, and, for $j\ge 1$, $p_{j}=p_{j}^{(*)}=(j+1)e^{\lambda_{0}+j\lambda_{1}-1}$ subject to the constraints 
\begin{align}
\sum_{k=0}^{\infty}k(k+1)e^{\lambda_{0}+k\lambda_{1}-1} = N \nonumber \\
\sum_{k=0}^{\infty}(k+1)e^{\lambda_{0}+k\lambda_{1}-1}=1 .
\end{align}

Solving the second constraint for $p_{0}^{(*)}$ gives $p_{0}^{(*)}=(1-e^{\lambda_{1}})^{2}$. Using this result in the first constraint, then solving the first constraint for $\lambda_{1}$ gives $\lambda_{1}^{(*)}=\ln {N\over N+2}$. Our expressions for $p_{0}^{(*)}$ give $e^{\lambda_{0}-1}=(1-e^{\lambda_{1}})^{2}$, from which it follows that $\lambda_{0}^{(*)}=1-2\ln {N+2 \over 2}$. Eliminating the critical Lagrange multipliers gives
\begin{align}
p_{0}^{(*)}&=\left( {2\over N+2} \right)^{2} \nonumber \\
p_{j}^{(*)}&=4(j+1){N^{j}\over (N+2)^{j+2}}.
\end{align}
Using these values in $\ket{\Phi}$ gives the state $\ket{\Phi_{N}}$ of maximal $C^{A}$ in Proposition \ref{prop:prop3}.  \hspace{.5cm}
$\blacksquare$

\section{\label{sec:pcsanal}Scaling of entropy of coherence for $\Psi(\theta = \pi/4,m=N/2)_{N}$}

To derive an analytical lower bound for the asymptotics of the entropy of coherence for the pair correlated state $\Psi(\pi/4,N/2)_{N}$, one can use the asymptotics for the central binomial coefficient. With $p(k):=\vert \langle k, N-k \vert \Psi(\pi/4,N/2)_{N} \rangle \vert^{2}$, one finds that as $N\rightarrow \infty$, $\min_{k}-p(k)\log_{2}p(k)\sim {4\over \pi N}(\log_{2}N + \log_{2}{\pi\over 4})$ and the minimum is achieved at $k=N/2$. Since $-x\log_{2}x$ is an increasing function on $(0,e^{-1})$ and the Fock state amplitudes of $\Psi(\pi/4,N/2)_{N}$ are increasing in modulus away from $\ket{{N\over 2},{N\over 2}}$, we have
\begin{eqnarray}
\sum_{k=0}^{N}-p(k)\log_{2}p(k)& \ge & \left( {N\over 2}+1\right) \min_{k}-p(k)\log_{2}p(k) \nonumber \\&\sim& {4 \left( {N\over 2}+1\right) \over \pi N}\left( \log_{2}N + \log_{2}{\pi\over 4}\right) \nonumber \\
&{}&{}
\end{eqnarray}
because $\Psi(\pi/4,N/2)_{N}$ has only ${N\over 2}+1$ nonvanishing amplitudes. Therefore, the entropy of coherence of $\Psi(\pi/4,N/2)_{N}$ relative to $\Delta^{A}_{2}$ is at least $O\left( \log_{2}N^{2\over \pi}\right)$.

\section{\label{sec:entmaxanal}Entropy of coherence for $\Phi_{N}$}

For $\Phi_{N}$, the joint probability density $p_{X,Y}(x,y)$ on the subset $K \subset \mathbb{Z}^{\times 2}$ defined by $K=\lbrace (x,y): \, x\ge 0 \, , \, 0\le y \le x \rbrace$ is given by $p_{X,Y}(x,y)=\left( { 2\over N+2 } \right)^{2} \left( {N\over N+2} \right)^{x}$, independent of $y$. It follows that
\begin{eqnarray}
C^{A}(\Phi_{N})&=& -\sum_{(x,y)\in K} p_{X,Y}(x,y) \log_{2}p_{X,Y}(x,y) \nonumber \\
&=& \sum_{(x,y)\in K} \left( { 2\over N+2 } \right)^{2} \left( {N\over N+2} \right)^{x} \log_{2} \left( { 2\over N+2 } \right)^{2} \left( {N\over N+2} \right)^{x} \nonumber \\
&=& \sum_{x=0}^{\infty} \left( { 2\over N+2 } \right)^{2} \left( {N\over N+2} \right)^{x}(x+1) \log_{2} \left( { 2\over N+2 } \right)^{2} \left( {N\over N+2} \right)^{x} \nonumber \\
&=&  {2N\over N+2} \log_{2}{N+2\over 2} +{N\over N+2}\log_{2} {N+2\over N}  + {4\over N+2}\log_{2}{N+2\over 2} \nonumber \\ &+& (N+1)\log_{2}{N+2\over N} \nonumber \\
&\sim & \log_{2}N^{2}.
\end{eqnarray}
Since $\log_{2}(2N+1)(N+1) \sim \log_{2}N^{2}$, (\ref{eqn:rmaxcoh}) shows that the asymptotically optimal distillation rate of $\widetilde{\text{MC}}_{N}$ from $\Phi_{N}$ is 1.

The state $\widetilde{\text{MC}}_{N}$ also exhibits large coherence in QCRTF$^{(A)}$, although it is not a state of maximal coherence in $D(\mathcal{F}_{2})$.
\begin{eqnarray}
C^{A}( \widetilde{\text{MC}}_{N} )&=&  \sum_{x=0}^{2N}\sum_{y=0}^{x} {1\over (2N+1)(x+1)}\log_{2} (2N+1)(x+1) \nonumber \\
&=&  {1\over 2N+1}\log_{2}2N+1+ \log_{2}2N+1 \nonumber \\ &+& {1\over 2N+1}\log_{2}2N+1!
\end{eqnarray}

\end{document}